\documentstyle[aps,prd]{revtex}
\font\fontA=cmss17
\def\be{\begin{equation}}
\def\ee{\end{equation}}
\def\bea{\begin{eqnarray}}
\def\nn{\nonumber}
\def\eea{\end{eqnarray}}

\newcommand{\pp}{\partial}

\newcommand{\lmk}{\left(}
\newcommand{\rmk}{\right)}
\newcommand{\lkk}{\left[}
\newcommand{\rkk}{\right]}

\rightline {OCHA-PP-92}
\rightline {\today}
\begin{document}
\thispagestyle{empty}
\begin{center}
{\fontA 
Dissipative Quantum Tunneling from  Quantum Potential Approach \\
--- Back Reaction of Particle Creation ---}
\par
\vskip1truecm
{\sc Fumiaki~SHIBATA}
\footnote{shibata@phys.ocha.ac.jp}
~~~{\sc Masahiro~MORIKAWA}
\footnote{hiro@phys.ocha.ac.jp}\\
\sl{ Department of Physics, Ochanomizu University \\[2mm]
1-1, Otsuka 2, Bunkyo-ku, Tokyo 112 JAPAN  }
\vskip1truecm
{\sc Tetsuya SHIROMIZU}
\footnote{siromizu@utaphp6.phys.s.u-tokyo.ac.jp}\\
\sl{Department of Physics, The University of Tokyo, Tokyo 113 JAPAN \\
and \\
Research Center for the Early Universe(RESCEU), \\ The University of Tokyo, 
Tokyo 113 JAPAN}
\par
\vskip1truecm
{\sc Masahide YAMAGUCHI}
\footnote{gucci@utaphp6.phys.s.u-tokyo.ac.jp}\\
\sl{Department of Physics, The University of Tokyo, Tokyo 113 JAPAN}
\end{center}
\vskip3cm
\begin{abstract} 
Back reaction of the particle creation on the quantum tunneling process is 
analyzed in real time formalism.   
We use quantum potential method in which whole quantum dynamics is exactly 
projected to a classical Hamilton-Jacobi equation with quantum corrections.  
We derive the reduction of the tunneling rate due to this particle creation 
effect.  
%\noindent{\bf pacs number: }
\end{abstract}
%\vfill
\vskip1cm
%\multicols{2}
%======================================%
%<<<<<<<<<<<< SECTION I  >>>>>>>>>>>>>>%
%======================================%
%\baselineskip25pt
\newpage
%\section{Introduction}
%    section I 
\par
Macroscopic quantum tunneling (MQT) is the fundamental dynamics underlying 
various fields in Physics such as the cosmological phase transitions in the very 
early Universe \cite{rubakov}\cite{vachaspati}, the macroscopic nucleation in 
$^3He-^4He$ liquid mixtures \cite{satoh},  and phase transitions in nuclear 
physics and in statistical mechanics.  
MQT process necessarily accompanies many degrees of freedom and is therefore 
dissipative.  
This dissipative MQT process is mainly analyzed so far in the imaginary time 
formalism utilizing the instanton method \cite{caldeira}.   

\par
However, this imaginary time formalism is not convenient when we study the back 
reaction of particle creation on the tunneling rate.  
During the system tunnels, it necessarily activates the environmental degrees of 
freedom which couple to the system.  
The energy of the system is dissipated in this process and the tunneling rate is 
thought to be reduced.  
If we force to use the imaginary time formalism, we encounter non-Unitary 
evolution equation for the wave function which cannot be normalizable. \cite{rubakov}  
Furthermore in the imaginary time formalism in general, qualitatively different effects such as dissipation and fluctuation cannot be separated in the tunneling dynamics.  

\par
Therefore the real time formalism for the quantum tunneling process is urgently 
necessary.  
We explore in this paper such possibility utilizing the quantum potential method 
\cite{bohm}.  
In this method, the quantum mechanics is exactly projected on the classical 
mechanics with quantum modification of the potential term.  
This complete one-to-one correspondence between the quantum mechanics and the 
classical mechanics is the starting point of our real time analysis.  
The small probability of quantum tunneling, in this quantum potential method, is 
not associated with the instanton or the saddle point in imaginary time but with 
a slow-rolling particle on the modified potential in real time.  
Therefore the dynamics of particle creation can be analyzed with the Unitary 
evolution of wave functions.   

%\par
%The rest of this paper is organized as follows. 
%In Sec. 2, we summarize the quantum potential method. 
%In Sec. 3, we apply the quantum potential method to the quantum tunneling 
%process.  
%In Sec. 4, we study the back reaction of the particle creation on the quantum 
%tunneling.   
%In Sec. 5, we calculate the particle creation and the resulting modification of 
%the tunneling rate.    
%In Sec. 6,  summarize our work.  

%======================================%
%<<<<<<<<<<< SECTION II  >>>>>>>>>>>>>>%
%======================================%

%\section{Quantum potential method}
%    section II
\par
Now we explain the quantum potential method, which was first proposed by 
Bohm\cite{bohm}. 
He attempted to construct a new interpretation of quantum mechanics(ontological 
basis) and applied it to the measurement theory of quantum mechanics. 
In this paper we skip the conceptual aspects of this method and directly apply 
it to the perturbation calculation for the modified tunneling rate. 
\par
We start from the one-dimensional quantum system which is described by the 
Schr\"odinger equation  
%========<Equation>========%
%
\be
i \hbar \frac{\pp}{\pp t} \Psi
= 
\lkk - \frac{\hbar^2}{2M}\frac{\pp^2}{ \pp x^2} +V(x) \rkk    \Psi .
\label{schroedinger}
\ee
%
%==========================%
The polar decomposition of the wave function  
$\Psi=R\exp(iW / \hbar)$
,with $R$ and $W$ being real functions of $x$ and $t$, yields an another form of 
the Schr\"odinger equation:
%========<Equation>========%
%
\be
\frac{W'^2}{2m}+V+V_Q=-{\dot W} 
\label{HJ}
\ee
%
%==========================%
and
%========<Equation>========%
%
\begin{equation}
2W{\dot R}+W''R+2W'R'=0,
\label{continuity}
\end{equation}
%
%==========================%
where $V_Q$ is the quantum potential defined by 
%========<Equation>========%
%
\begin{equation}
V_Q=-\frac{\hbar^2}{2M} \frac{R''}{R}.  
\end{equation}
%
%==========================%
The over dot and the prime respectively denote the time and spatial derivatives.  
The latter equation Eq.(\ref{continuity}) simply represents the conservation 
${\rm div}J(t,x)=0$
of the probability flow $J(t,x)$ defined by
\be
J(t,x)=(R(t,x)^2, W(t,x)'R (t,x)^2/M).
\label{probcurrent}
\ee
The former equation Eq.(\ref{HJ}) is the Hamilton-Jacobi equation in classical 
mechanics if we regard $W(t,x)$ as the action function and the total potential 
$V_{tot}$ as  $V_{tot}=V+V_Q$.  
Thus there exists a corresponding classical system for any quantum system and 
this correspondence is exact; it is not any approximation such as WKB.  
This corresponding classical mechanics (effective classical system) is generally 
different from the underlying classical system (original classical system) which 
is supposed to be quantized to yield the quantum mechanics described by 
Eq.(\ref{schroedinger}).  
The whole quantum nature is now concentrated on the quantum potential $V_Q$.  
This correspondence can be extended to systems of many degrees of freedom and to 
relativistic systems.  

\par
This quantum potential method is particularly useful for stationary states.   
In this case, the variable $R$ becomes independent of time, and 
$\dot W= - E$.  
Therefore $V_Q$ becomes a time-independent potential.  
Moreover $E-V_{\rm tot}=W'^2/(2M)>0$ and therefore there appear no classically 
forbidden region in the configuration space of the effective classical system.  
This fact is particularly interesting when the method is applied to the quantum 
tunneling process in which a particle in the original classical mechanics cannot 
go over the tunneling region in the configuration space.  
On the other hand, the quantum fluctuations modify the potential so that a 
particle in the effective classical system can go over the tunneling region.   
In the effective classical system, the tunneling process is described by a 
slowly rolling particle on the modified potential at the tunneling region in 
real time.  
This real time formalism should be compared with the ordinary imaginary time 
formalism using the saddle point method in the path integral (instanton method).  

%======================================%
%<<<<<<<<<<< SECTION III  >>>>>>>>>>>>>%
%======================================%

%\section{Quantum tunneling in the quantum potential method}
%   section III
Now we apply the quantum potential method to the quantum tunneling dynamics.  
We consider a stationary problem for using the advantage of the quantum 
potential method.  
First we study a simple one-dimensional model with a rectangular potential, 
which is given by 
%========<Equation>========%
%
\bea
V=
\left\{
\begin{array}{rcl}
0     & \mbox{for $x<0$}     & \mbox{(region I)}\\
V_0 & \mbox{for $0<x<a$} & \mbox{(region II)}\\
0     & \mbox{for $a<x$}     & \mbox{(region III)}
\end{array}
\right., 
\label{rectangularpot}
\eea
%
%==========================%
with $V_0$ being a positive constant.   
The tunneling wave function in each region is given by 
%========<Equation>========%
%
\be
\phi_I(x)=A e^{i k x}+ B e^{-i k x},  ~~
\phi_{II}(x)=F e^{-\beta x}+ G e^{\beta x},  ~~
\phi_{III}(x)=C e^{i k x},
\label{rectwf}
\ee
%
%==========================%
with
$k=\sqrt{2ME}/\hbar,  ~~~ \beta=\sqrt{2M(V_0-E)}/\hbar$.   
Smoothness conditions of the wave function at the edges of the potential ($0$ 
and $a$) yield
%========<Equation>========%
%
\be
\left( {\matrix{F\cr
G\cr
}} \right)
=\frac{C e^{i k a}}{2 \beta}
\left( {\matrix{\lambda_- e^{\beta a} \cr
\lambda_+ e^{- \beta a}\cr
}} \right), ~~
\left( {\matrix{A\cr
B\cr
}} \right)
=\frac{-C e^{i k a}}{4 i k \beta}
\left( {\matrix{\lambda_-^2 e^{\beta a}-\lambda_+^2 e^{-\beta a} \cr
\lambda_-\lambda_+(e^{-\beta a}-e^{\beta a})\cr
}} \right),
\label{ABCFG}
\ee
%
%==========================%
where 
$\lambda_{\pm}=\beta \pm i k$.  
This wave function yields the quantum potential in each region as is shown in 
Figs. 1a and 1b.  
%%%%%%%%%%%%%%%%%%%%%%
\begin{center}
{\bf \Huge Figs. 1a, 1b}
\end{center}
%%%%%%%%%%%%%%%%%%%%%%
Especially the total potential in region II is given by 
%========<Equation>========%
%
\bea
E-V_{\rm tot}&=&\frac{\hbar^2 \beta^2}{2M}
\frac{4k^2\beta^2}{[(k^2 +\beta^2)\cosh(2\beta(a-x))+(\beta^2-k^2)]^2}.  
\label{totpot}
\eea
%
%==========================%
The exponentially small tunneling probability when $\beta a \gg 1$ is reflected 
in the fact that in region II, the kinetic energy $E-V_{\rm tot}$ of the 
classical particle is exponentially small.  
Actually it is possible to calculate the classical ``rolling time'' over the 
tunneling region II:
%========<Equation>========%
%
\bea
t_{\rm roll}&=&\int dt = \int_0^a\frac{dx}{v} \nn\\
&=&\frac{M}{4 \hbar k \beta^2}
[\frac{k^2+\beta^2}{\beta}\sinh(2\beta a)+2a(\beta^2-k^2)].
\label{rollingtime}
\eea
%
%==========================%
On the other hand the tunneling probability is given by 
%========<Equation>========%
%
\be
P=\frac{|C|^2}{|A|^2} 
=\frac{4 k^2 \beta^2}
{(k^2+\beta^2)^2\cosh^2(\beta a)-(\beta^2-k^2)^2}.
\label{tunnelprob}
\ee
%
%==========================%
When the potential barrier is high ($\beta a \gg 1$), the tunneling probability 
is proportional to the inverse of the rolling time.  
%========<Equation>========%
%
\be
P \cdot t_{\rm roll} \approx 
\frac{\hbar \sqrt{E}}{4  V_0 \sqrt{V_0-E}}.
\label{rolltimeandprob}
\ee
%
%==========================%
Therefore the rolling time can be an another measure of the quantum tunneling.  
Actually this rolling time is known as the ``traversal time'' \cite{landauer}.  

\par
For a general form of one-dimensional potential, we can use the WKB 
approximation with some modification near the turning points. 
We use the connection formula at the turning point $x_0$ where the potential 
$V(x)$ increases toward right ($V'(x_0)>0$):
%========<Equation>========%
%
\bea
\begin{array}{rl}
\frac{1}{\sqrt{|p|}}  \left[ \exp{(\int_{x_0}^x |p| dx)} \pm 
                                     \frac{i}{2}\exp{(-\int_{x_0}^x |p| dx)}
                                    \right]
    & \mbox{in $E<V$ side}\\ 
\Longleftrightarrow ~~~
\frac{1}{\sqrt{p}}\exp{\pm i (\int_x^{x_0} p dx +\frac{\pi}{4})}
    & \mbox{in $E>V$ side},
\end{array}
\eea
%
%==========================%
where $p=\sqrt{2M(E-V(x))}/\hbar$.  
Similar connection form at the turning point with $V'(x_0)<0$ holds if we 
replace $\int_{x_0}^x \leftrightarrow \int_x^{x_0}$.  
It is important that we cannot drop the exponentially small term which is pure 
imaginary.  
This small but imaginary term yields non-vanishing kinetic energy of the rolling 
particle.  
The graph of the total potential is shown in Fig. 2.    
The total potential in the region near the turning points are estimated by using 
the airy functions which appear when we linearize the potential there.  
%%%%%%%%%%%%%%%%%%%%%%
\begin{center}
{\bf \Huge Fig. 2}
\end{center}
%%%%%%%%%%%%%%%%%%%%%%
%======================================%
%<<<<<<<<<<< SECTION IV  >>>>>>>>>>>>>>%
%======================================%

%\section{The back reaction of the particle creation}
%   section IV
\par
Before we calculate the precise particle creation effect during the quantum 
tunneling of the system, we clarify the general back reaction effect.  
The Hamiltonian of the total system we consider is 
%========<Equation>========%
%
\bea
H&=&H_s(x)+H_e(x,\{y_n\}),\nn\\
H_s(x)&=&\frac{p_x^2}{2M}+V(x),~~~
H_e(x,\{y_n\})=\sum_n \left[ 
              \frac{p_n^2}{2m_n}+
             \frac{m_n}{2}\omega_{n0}^2 y_n^2 + c~ x~ y_n^2
                                 \right].
\eea
%
%==========================%
We introduce the polar decomposition of the total wave function $\Phi$ which is 
assumed to have the factorized form,
%========<Equation>========%
%
\bea
\Phi(x,\{y_n\})&=&\phi_s(x)\phi_e(x,\{y_n\}),\nn\\
\phi_s(x)&=&R_s(x)e^{iW_s(x)/\hbar}, \nn\\
\phi_e(x,\{y_n\})&=&\prod_n R_n(x,y_n)\exp(\sum_n iW_n(x,y_n)/\hbar).
\eea
%
%==========================%
Putting this decomposition into the total stationary Schr\"odinger equation for 
$\Phi(x,\{y_n\})$, we obtain
%========<Equation>========%
%
\bea
&&\frac{\lmk {W_s}'(x) +  \sum_n W_n' \rmk^2}{2M}  +  V(x)-\frac{\hbar^2}{2 M}\frac{(R_s\prod_n R_n ) ''}
{R_s \prod_n R_n}\nonumber \\
&& +  \sum_n \lkk \frac{(\partial_{y_n}W_n)^2}{2m_n}+\frac{1}{2} m_n \omega_{n0}^2y_n^2+cxy_n^2 
-\frac{\hbar^2}{2m_n}\frac{\partial_{y_n}^2R_n}{R_n} \rkk =  E. 
\label{OGHJ}
\eea
%
%==========================%
where the prime denotes the derivative with respect to $x$.

\par
In order to study the back reaction of the particle creation of the environment 
on the quantum tunneling of the system, we use the quantum potential method as a perturbation from the pure tunneling without particle creation.  
We suppose that the tunneling dynamics of the system alone without back reaction is already known with the system wave function 
$R_s^0 \exp(i W_s^0/\hbar)$.  
Then there is an associated classical motion $\bar x(t)$ derived from Eq.(\ref{HJ}).  
We turn on the coupling between the tunneling system and the environment. 
If this tunneling degrees of freedom is macroscopic and semi-classical, then we 
can define ``time'' for the environment by using this solution:
$\partial/\partial t=({W_s^0}'(\bar x)/M) \partial/\partial \bar x$.  
Then the wave function of the environment obeys the time-dependent Schr\"odinger equation with respect to this time
%========<Equation>========%
%
\be
i \hbar \frac{\partial \phi_e}{\partial t}=H_e(\bar{x}(t),\{y_n\}) \phi_e
\label{evolution}
\ee
%
%==========================%
in the order $O(\hbar)$. 
A non-trivial time dependence enters from the tunneling solution $\bar x(t)$ which causes particle creation.  
\par
Aided by Eq.(\ref{evolution}), the HJ equation Eq.(\ref{OGHJ}) becomes simpler, 
%========<Equation>========%
%
\bea
&&\frac{{W_s}'(x)^2}{2M}+V(x)-\frac{\hbar^2}{2 M}\frac{R_s ''}{R_s}-E+\frac{1}{M} 
({W_s}'-{W_s^0}')\sum_n W_n'+ \nn\\
&&+\frac{(\sum_n W_n')^2}{2M}-\frac{\hbar^2}{2M}(2\frac{R_s'}{R_s}
\sum_n \frac{R_n'}{R_n}+\sum_n \frac{R_n''}{R_n}+
\sum_{m \ne n}\frac{ R_m'}{ R_m}\frac{ R_n'}{ R_n})=0.  
\label{GHJ}
\eea
%
%==========================%
\par
We suppose that the environment state is well approximated by the Gaussian 
state.  
This is guaranteed by the Gaussian initial state and the semi-free evolution of 
the environment.  
A general Gaussian state is expressed by the wave function of the form
%========<Equation>========%
%
\be
\phi_e(t,\{y_n\})=\prod_n \pi^{-1/4}\alpha_n(t)^{1/2} 
\exp[\frac{1}{\hbar}(-\alpha_n(t)^2+i \beta_n(t))\frac{y_n^2}{2}].
\ee  
%
%==========================%
where the real variables $\alpha_n(t)$ and $\beta_n(t)$ obey the equations of 
motion:
%========<Equation>========%
%
\be
\dot \alpha_n=-\frac{\alpha_n\beta_n }{m_n}, ~~~ 
\dot \beta_n=\frac{\alpha_n^4}{m_n}-\frac{\beta_n^2}{m_n}-m_n \omega_n(t)^2.  
\label{eqalphabeta}
\ee
%
%==========================%
Time dependent frequency for the mode $n$,  $\omega_n(t)=\sqrt{\omega_{n0}^2+2c\bar 
x(t)/m_n}$ comes from the motion of the tunneling solution $\bar x$.  

We are interested in the effective motion of the system ($x$) and therefor should integrate out the environmental degrees of freedom. 
This integration is simply a quantum averaging:
%========<Equation>========%
%
\bea
\langle F[\lbrace y_n \rbrace ] \rangle = \frac{\int\prod_n  dy_n |\phi_e|^2F[\lbrace y_n \rbrace ]
 }{\int \prod_n  dy_n |\phi_e|^2}. 
\eea
%
%==========================%
After this average, the HJ equation finally becomes
%========<Equation>========%
%
\bea
\frac{{W_s}'(x)^2}{2M}+V(x)-\frac{\hbar^2}{2 M}\frac{R_s ''}{R_s}-E+\frac{\hbar}{M} 
({W_s}'-{W_s^0}')\sum_n \frac{\beta_n'}{4 \alpha_n^2}
&+& \nn\\
+\frac{\hbar^2}{2M}(\frac{3}{16}\sum_n \frac{\beta_n'^2}{\alpha_n^4}+
                        \frac{1}{16}\sum_{m \neq n} \frac{\beta_n'}{\alpha_n^2}  
\frac{\beta_m'}{\alpha_m^2})
+\frac{\hbar^2}{4 M}\sum_n \frac{{\alpha_n'}^2}{\alpha_n^2}&=&0.  
\label{AGHJ}
\eea
%
%==========================%
There are two classes of back reaction terms.  
The first class terms, the second line of Eq.(\ref{AGHJ}), are positive definite 
and they contribute to increase the classical potential.  
Therefore they systematically reduce the tunneling rate of the system.  
The second class term, the last term in the first line of Eq.(\ref{AGHJ}), 
does not have definite signature and may increase or reduce the tunneling rate.  
It will be interesting that this term is linear in the momentum $W_s'$ which 
reminds us the form of friction.

%======================================%
%<<<<<<<<<<<< SECTION V  >>>>>>>>>>>>>>%
%======================================%

%\section{Particle creation and its back reactions}
\par
We now concretely calculate the particle creation and its back reaction on the 
tunneling rate of the system.  
First we study the model of rectangular potential.  
Though the analytic form of the tunneling solution is known 
from Eq.(\ref{totpot}), it would be unnecessarily complicated and less general 
to calculate the normal modes for the time dependent frequencies.  
Therefore we approximate the classical solution $\bar x(t)$ in the 
hyperbolic-tangent form: 
$\bar x(t)=a(1+\tanh(\rho t))$ with $\rho=\hbar k/(aM)$. 
This form most faithfully represents the dynamics near the right turning point 
where the particle creation is thought to be maximum\footnote{
This type of bold approximation is inevitable for analytical calculations. }.  
We would like to solve the evolution equation Eq.(\ref{eqalphabeta}) with this 
time dependent background.  
Changing the variable from $\alpha_n, \beta_n$ to $\xi_n$ by 
$\ln(\xi_n(t)\dot)=(\beta_n+i \alpha_n^2)/m_n \hbar$, 
Eq.(\ref{eqalphabeta}) reduces to the form
$\ddot\xi_n(t)+\omega_n(t)^2 \xi_n(t)=0$.  
The solution of this equation can be represented by the hyper-geometric 
functions.  
Especially the solution which asymptotically approaches to the vacuum for $t 
\rightarrow -\infty$ is given by
%========<Equation>========%
%
\bea
\xi_n(t)&=&(2 \omega_{n0})^{-1/2}\exp(i \omega_{n+}t+
            i \omega_{n-}\ln(2\cosh(\rho t))/\rho) \times \nn\\
          & &~~\times~~  _2F_1[1-\frac{i \omega_{n-}}{\rho}, -\frac{i 
\omega_{n-}}{\rho};
            1+\frac{i \omega_{n0}}{\rho};\frac{1}{2}(1+\tanh(\rho t))], 
\label{xi}
\eea
%
%==========================%
where $\omega_{n\pm}=(\sqrt{\omega_{n0}^2+(4ca)/m_n}\pm \omega_{n0})/2$.  
Then the typical back reaction terms in Eq.(\ref{AGHJ}) are all calculated from 
this solution $\xi_n(t)$:
%========<Equation>========%
%
\be
Q_{n1} \equiv \frac{\beta_n'}{\alpha_n^2}=
    \frac{\Re (\ln \xi_n \ddot)}{\dot{\bar x}~\Im (\ln \xi_n \dot)}, ~~~
Q_{n2} \equiv \frac{{\alpha_n'} ^2}{\alpha_n^2}=
     \left(\frac{\Im (\ln \xi_n \ddot)}{2\dot{\bar x}~\Im (\ln \xi_n 
\dot)}\right)^2.  
\ee
%
%==========================%
Since the total back reaction is simply a superposition of each single mode, we 
concentrate on the back reaction effect from a single mode in the following.  
The result of the numerical calculation for a single mode $n$ is given in Fig. 
3.  
Analytically the special case of $\rho=\omega_{n0}$  can be easily calculated in 
the series of $\epsilon=2c a /(m \rho^2)$,
%========<Equation>========%
%
\be
Q_1|_{x=a}=-0.272029 \frac{\epsilon}{a}+O(\epsilon^2),~~~
Q_2|_{x=a}=0.14538 \frac{\epsilon^2}{2 a^2} +O(\epsilon^3).  
\ee
%
%==========================%
\par
Now we evaluate the modified tunneling rate due to the particle creation of a 
single mode.  
The total tunneling rate can be derived by adding all the single mode effects.  
The effective Hamiltonian for the system is read from Eq.(\ref{AGHJ}) excluding the quantum potential,  
%========<Equation>========%
%
\be
H_{eff}=\frac{p^2}{2M}+V(x)+\frac{\hbar Q_1}{4M}(p-p_0)+
\frac{3 \hbar^2 Q_1^2}{32M}+\frac{\hbar^2 Q_2}{4M}, 
\ee
%
%==========================%
where $p_0=W_{0s}'$.  
Then the corresponding Hamilton equation of motion can be brought into a simple form 
%========<Equation>========%
%
\bea
&&M \ddot x + V_{eff}'=0, ~~{\rm with}~~ \nn\\
&&V_{eff}=V(x)+\frac{2 \hbar^2 Q_1^2}{32M}+\frac{\hbar^2 Q_2}{4M}
-\int_{0}^{x}\frac{\hbar Q_1'(x)}{4M}p_0(x)dx.
\label{effeqmotion}
\eea
%
%==========================%
  
\par
Both classes of back reaction mentioned just after Eq.(\ref{AGHJ}) can be expressed in terms of the effective potential $V_{eff}$ which is modified from the original potential $V(x)$.  
In the present case of particle creation from the vacuum, we numerically find $Q_1<0$ as well as $Q_1'<0$ (Fig. 3). 
%%%%%%%%%%%%%%%%%%%%%%
\begin{center}
{\bf \Huge Fig. 3}
\end{center}
%%%%%%%%%%%%%%%%%%%%%%
Therefore we conclude $V_{eff}>V(x)$ and  the modification of the potential systematically reduces the tunneling rate.  
This is consistent with the fact that both classes reduce the momentum of the 
classical particle and increase the tunneling time defined in 
Eq.(\ref{rollingtime}).    
\par
The systematic reduction of the tunneling rate comes from the potential shift 
$\Delta V= V_{eff}-V$.  
%========<Equation>========%
%
\be
P=P(^{\rm no ~ particle}_{\rm creation})\times
\left[1+(\frac{1}{\beta^2}-\frac{2}{\beta^2+k^2})\frac{2M \Delta V}{\hbar^2}
\right]
\exp(-\frac{2M a \Delta V}{\beta \hbar^2}).  
\label{redtunnelprob}
\ee
%
%==========================%
\par
In the similar way, but using WKB approximation as well as the modification of 
it near the turning points, we can derive the modified tunneling rate for a 
general potential of the form in Fig 2.  
If we adopt the same strategy to approximate the classical motion by the 
hyperbolic-tangent form, the parameter $\rho$ should be given by
$\rho=(3^{5/6}\Gamma(2/3)/2\Gamma(1/3))(\hbar \beta^{1/3}/Ma)$, where
$\beta=-V'(a)$ with $x=a$ being the right turning point.

%======================================%
%<<<<<<<<<<<< SECTION VI  >>>>>>>>>>>>>%
%======================================%

%\section{Conclusions and Discussions}
\par
We now summarize our work.  
We first introduced the quantum potential method in which the quantum mechanics is exactly mapped to some classical mechanics (effective classical system) which is generally different from the original classical system.  
When this method is applied to the quantum tunneling process, the effective 
classical particle can slowly roll down the tunneling region in the configuration 
space.  
This classical motion in real time corresponds to the instanton in the imaginary 
time formalism.  
This rolling motion of the tunneling system induces the particle creation of the 
environmental degrees of freedom which were initially in the ground state.  
As a back reaction of this dissipation of energy, the effective potential is 
increased and therefore the tunneling rate of the system is reduced.  
\par
Several comments on our future work are in order.  
\par
This reduction of the tunneling rate is associated with the initial ground state 
of the environment.  
It would be interesting to study the particle creation effect for the initial 
excited state of the environment.  
In this case, the induced particle creation effect would further reduce the 
tunneling rate while the energy transfer from the environment to the system 
would reduce the rate.  
\par
We started from the classical solution $\bar{x}(t)$ of the effective classical 
system, and calculated the modification of the effective potential.  
Finding a new solution for this effective potential, we can repeat 
this quantum potential method.  
Then we may be able to find a series of solution which asymptotically approaches 
the true wave function including the full back reaction of particle creation.  
This iteration method would be another approximation paradigm which is free from 
any turning points since there is no classically forbidden region in the 
effective classical system. 
\par
We can directly apply our method to the quantum cosmology.  
Especially our method would reveal back reaction of the cosmological particle 
creation on the tunneling solution of the Wheeler-DeWitt equation.  
We will soon reported these work.  

%======================================%
%<<<<<<<<<<<< REFERENCES >>>>>>>>>>>>>>%
%======================================%

\par
\noindent
{\Large {\bf Figure captions}}
\begin{itemize}
\item[Fig. 1a: ] 
The rectangular potential $V$ (dashed graph), and the associated total potential 
$V_{tot}$ in the tunneling region (solid graph).   
We set the energy $E=2$(dotted graph), $a=1,V_0=4$ in the units $\hbar=1, M=1$.  
\item[Fig. 1b: ]
Same as Fig. 1a but in global view.  
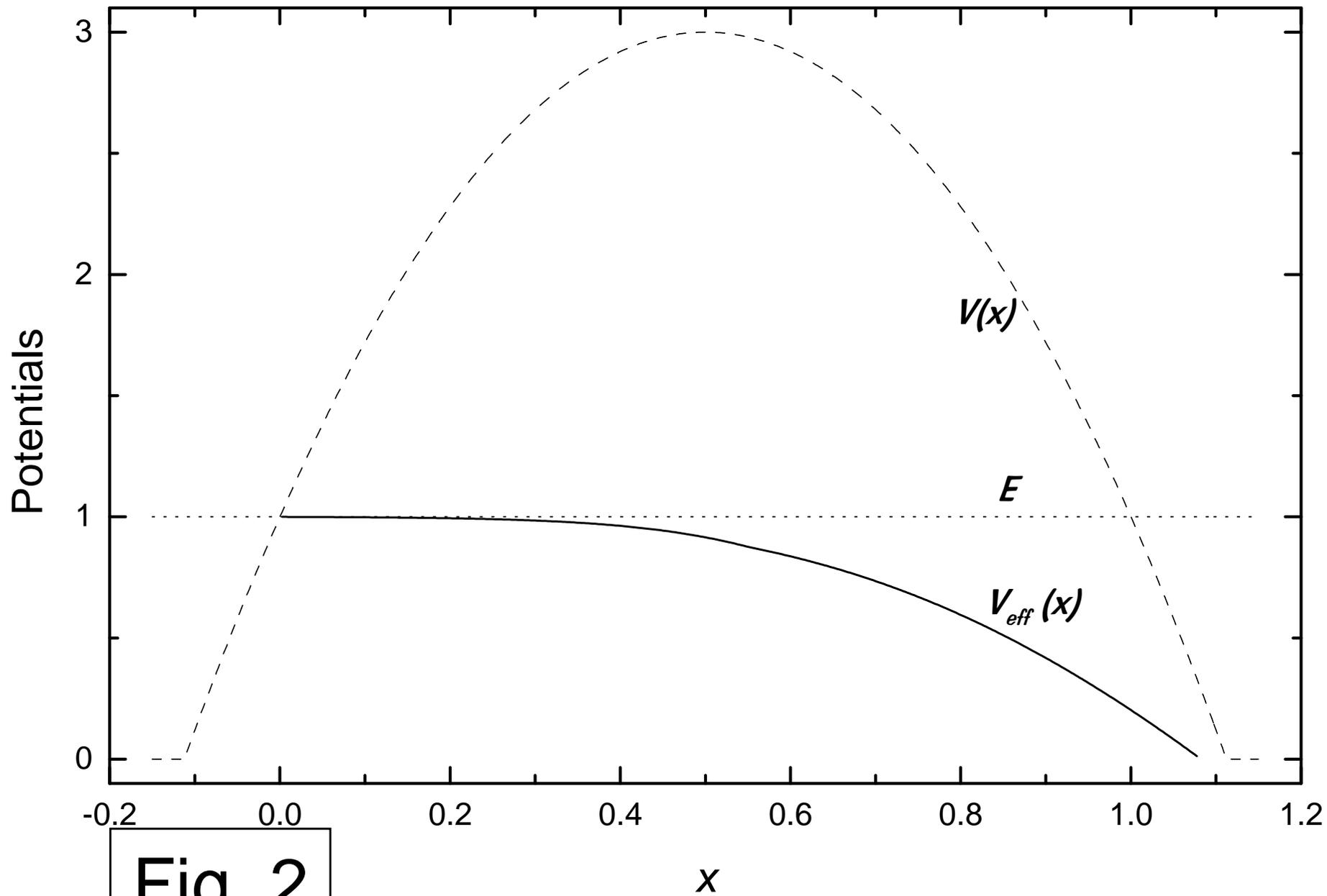
\item[Fig. 2: ]
The quadratic potential $V(x)=1-8x(x-1)$ (dashed graph) and the associated total potential $V_{tot}$ (solid graph) calculated from WKB approximation with modification by Airy function near the turning points.    
We set the energy $E=1$(dotted graph) in the units $\hbar=1, M=1$.
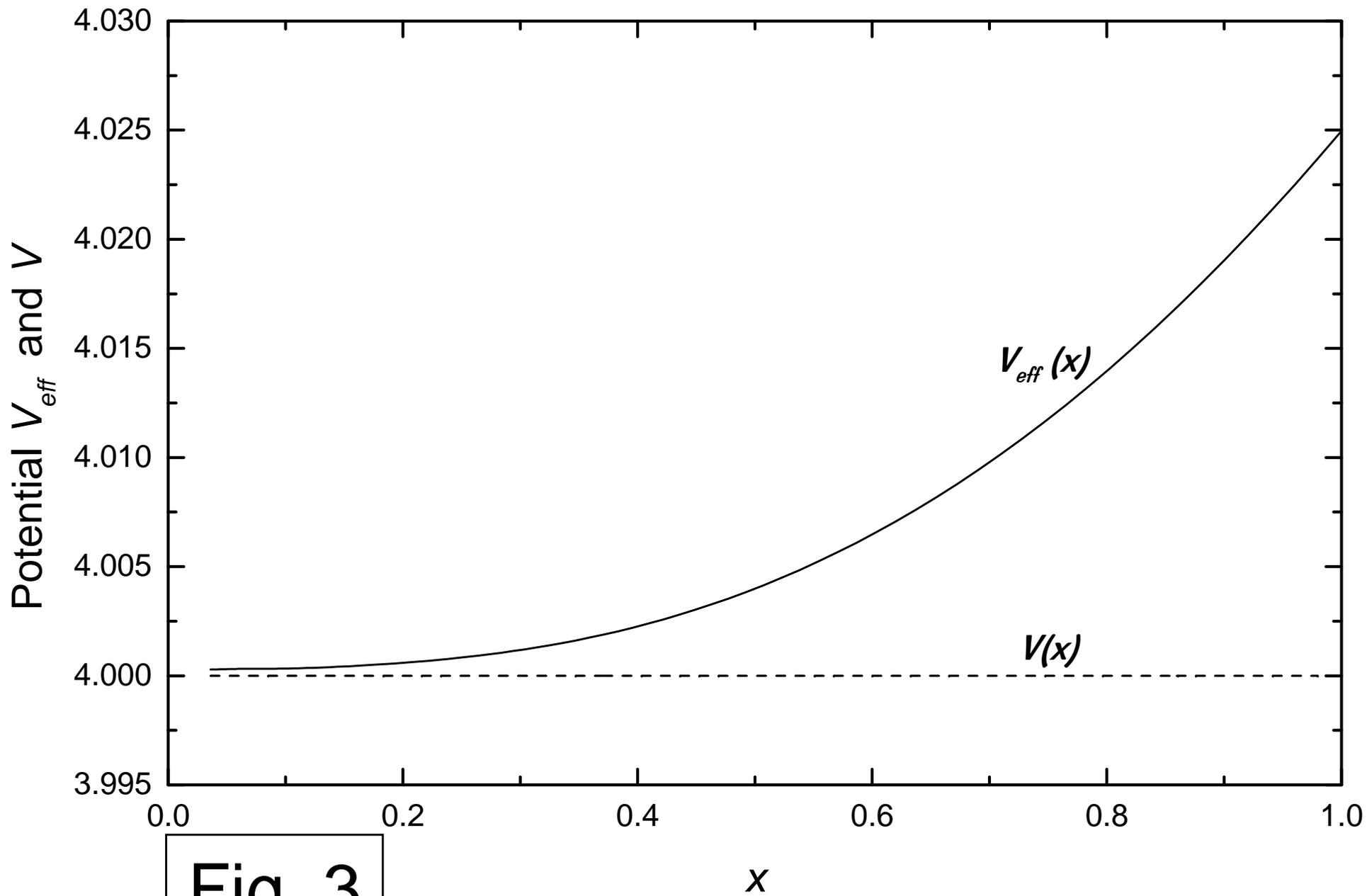
\item[Fig. 3: ]
The particle creation effect on the tunneling rate is fully represented by the effective potential $V_{\rm eff}(x)$ (solid graph) which is calculated for the rectangular potential model.  
We set the parameter as $E=2, a=1, V_0=4, m=1, \omega_0=1, c=0.15$ in the units $\hbar=1, M=1$.
\end{itemize}

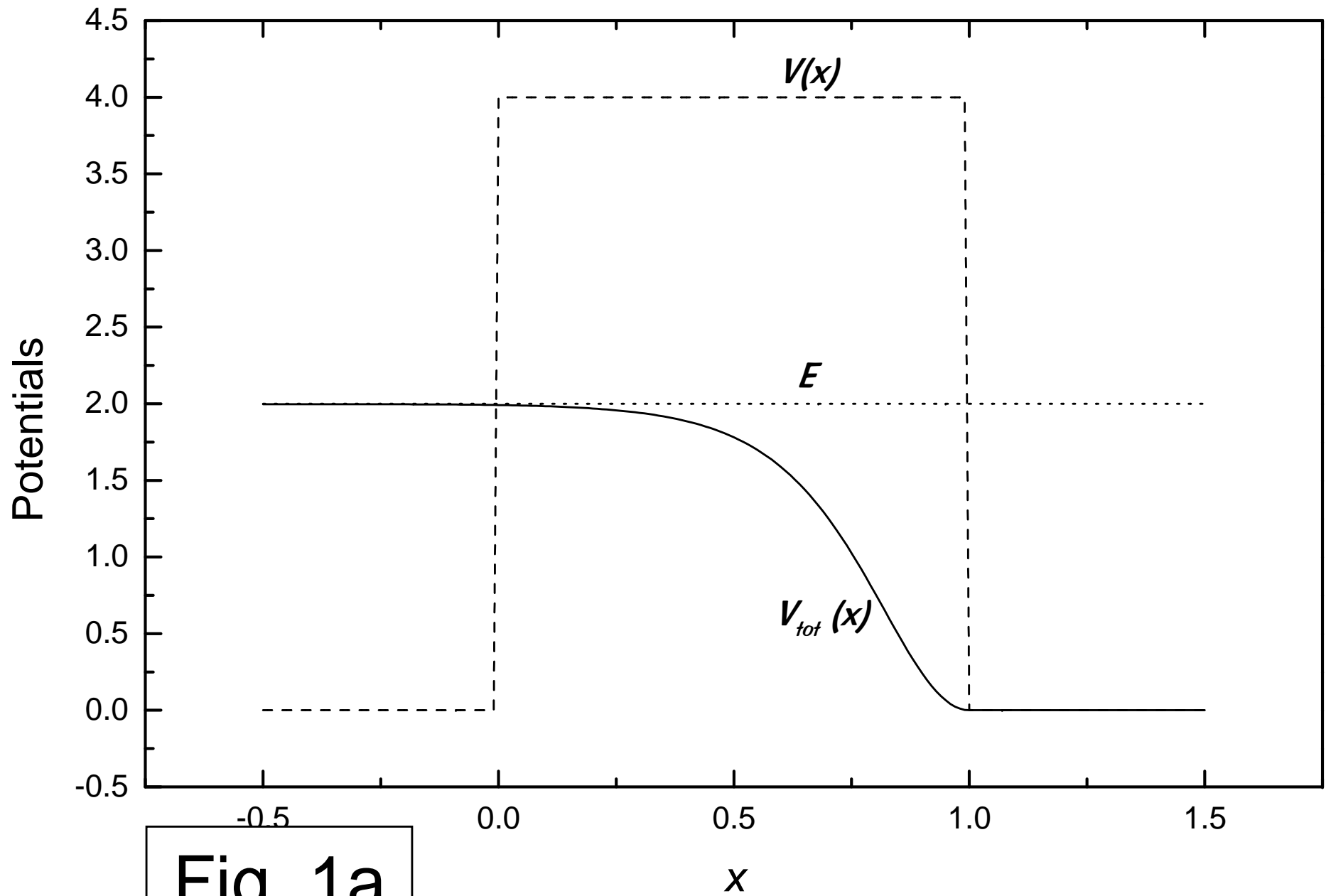
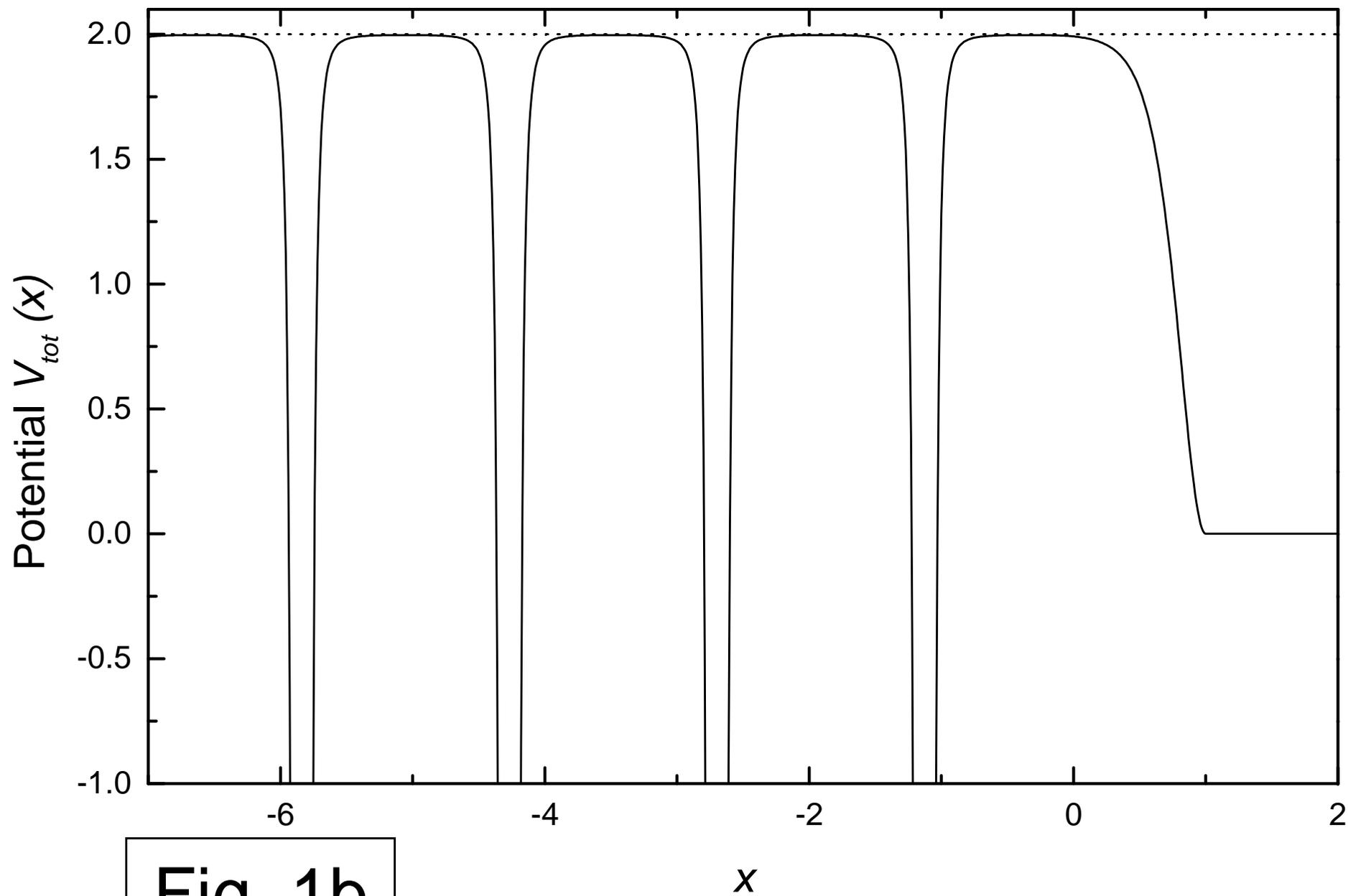
\end{document}